\documentstyle[12pt,aaspp]{article}
\def\1e{1E$1740.7-2942$}
\begin{document}

\title{Images of HCO$^+(1-0)$ Emission in a Molecular Cloud near
	1E1740.7$-$2942}

\author{J.A. Phillips}
\affil{Owens Valley Radio Observatory, Caltech 105-24, Pasadena, CA  91125}

\and
\author{T. Joseph W. Lazio}
\affil{Department of Astronomy and NAIC, Cornell University, Ithaca,
NY 14853-6801}

\begin{abstract}
We have observed the hard X-ray source \1e in the HCO$^+$(1-0) line
using the Owens Valley millimeter interferometer.  Previous
single dish observations have found HCO$^+$ emission coincident with the
location of the radio continuum hot spots of the radio source associated
with \1e.  Our higher resolution observations show a 15\arcsec\  offset
between the HCO$^+$ emission and the location of the radio hot spots.
We propose that the lack of emission results from a large ionization
rate, exceeding $10^{-15}$ s$^{-1}$, in the neighborhood of \1e.
\end{abstract}

\section{Introduction}

The Galactic center region is a luminous source of 511 keV X-ray line
emission resulting from electron-positron annihilation.  The line has two
components:  A diffuse and steady component resulting from
annihilation of positrons produced by novae and supernovae in the
interstellar medium, and a time-variable component, long believed to
originate in a black hole or neutron star, dubbed the ``Great Annihilator''
(\cite{lin88}).  The properties of the time-variable line indicate that the
annihilation region is small, $< 10^{18}$ cm, dense, $n_{\rm H} > 10^5$
cm$^{-3}$, and cold, $T < 5 \times 10^4$ K.  This annihilation environment
is very different from the high energy environment in which the positrons
are produced, e.g. an accretion disk near a black hole, and there is
considerable interest in identifying the source of the positrons.

The hard X-ray source \1e (Sunyaev et al. 1991) is a strong candidate for the
positron source.  The emerging picture of \1e is the following
(\cite{che94};\cite{mir94}): it is a compact object, probably a
stellar-mass black hole, on the edge of a molecular cloud (Mirabel et al.
1991; Bally \& Levanthal 1991). The high-energy emission arises from an
accretion disk with instabilities giving rise to pair production and
annihilation which in turn produces the excess above 300 keV.  The accretion
disk collimates some of the pairs to form jets which are visible as
bipolar radio lobes (Mirabel et al. 1992).  The pairs in the jets travel
about 1 pc before slowing and thermalizing in the cold, dense environment of
the molecular cloud where any remaining positrons are annihilated.

The details of this picture are controversial.  It is not clear whether the
source of accreted material is a stellar companion or perhaps the molecular
cloud itself (\cite{cam93}; \cite{che94}).  Another possibility is that the
radio lobes around \1e are extra-galactic (e.g., \cite{esp92}), but Mirabel
\& Rodriguez (1993) have recently detected changes in the flux density of
the radio lobes at the 3$\sigma$ level which suggest that the source is
closer than 17 kpc.

One way to establish the Galactic origin of \1e would be to prove a physical
connection between the X-ray source and the molecular cloud.  A drawback of
previous observations of the molecular cloud is that they were made with
single-dish  telescopes and have moderate to poor angular resolution. To
explore the relationship between \1e and the molecular cloud, we undertook
interferometric observations in the HCO$^+\,(1 \to 0)$ line.  HCO$^+$ is
believed to be a tracer of energetic outflows in molecular gas (e.g.,
\cite{gc92}), and we hoped to detect HCO$^+$ emission from the vicinity of
the radio lobes around \1e. In \S2 we describe the observations and \S3 we
discuss the images we obtained and their implications for the Great
Annihilator.

\section{Observations and Results}

We observed HCO$^+$(1-0) line emission from \1e using the Owens Valley Radio
Observatory millimeter interferometer.   The primary beam width of the 10.4m
OVRO antennas (65\arcsec\ FWHM) was only slightly larger than the 1\arcmin\
extent of the radio jets (Mirabel et al. 1992), so we used four overlapping
pointings to map the field around \1e.  We observed two of the four fields
during the period 1992 December - 1993 July when the array consisted of
four dishes with cryogenically-cooled SIS receivers.  The two pointings
($\alpha=17^{\rm h}\,40^{\rm m}\,42\fs5$,
$\delta=-29^\circ\,43\arcmin\,15\arcsec$ and
$\alpha=17^{\rm h}\,40^{\rm m}\,43\fs5$,
$\delta=-29^\circ\,43\arcmin\,40\arcsec$, epoch 1950.0)
were centered on the VLA jets and offset by about $\pm 15\arcsec$ from the
X-ray core.  We alternated 5 minute integrations between the two phase
centers, observing the nearby point source NRAO530 every 20 minutes to
monitor the phase behavior of the array.
The absolute flux scale was determined from observations
of Uranus and Neptune. We observed 3C273 ($S=26.2$ Jy) for at least 15
minutes per transit to obtain a high signal-to-noise passband calibration.
The total on-source integration time was 8 hours per field and the average
single-sideband system temperature was 640 K.

In 1993 September we observed two more pointings offset from the X-ray
source by $\approx 30\arcsec$ west and northwest, respectively; the
 phase centers were
$\alpha=17^{\rm h}\,40^{\rm m}\,44\fs0$,
$\delta=-29^\circ\,43\arcmin\,00\arcsec$ and
$\alpha=17^{\rm h}\,40^{\rm m}\,45\fs0$,
$\delta=-29^\circ\,43\arcmin\,25\arcsec$.
By 1993 September the interferometer had been upgraded  to five elements, but
otherwise the equipment and observing procedures were the same as before.
The total integration time per field was 2.4 hours and the mean
single-sideband system temperature was 480 K.

During the observations we operated a digital spectrometer divided into two
bands. A 32 MHz (108 km s$^{-1}$) band with 128 frequency channels was
centered on the HCO$^+$ transition frequency (89.189 GHz) and Doppler
shifted to an LSR velocity of $-140$ km s$^{-1}$. A second, lower-resolution
band with 128 MHz bandwidth (431 km s$^{-1}$) and 128 frequency channels was
Doppler shifted to $-40$ km s$^{-1}$.  The high-resolution band was centered
on a strong HCO$^+$ line observed by Mirabel et al. (1991) using the IRAM
30m telescope.  The low-resolution band covered the velocity range -200 to
+100 km s$^{-1}$ where other molecular species have been detected in the
direction of \1e (\cite{mir91}).

We calibrated the data using software specific to the OVRO millimeter
interferometer and imaged each of the four fields
using AIPS.  Although the data for the first two and the second two fields
were obtained with different numbers of telescopes and array configurations,
the u-v coverage was very similar for the four data sets.  Natural weighting
of the visibility data gave a $17\arcsec \times 7\arcsec$ CLEAN beam at
position angle
6\deg\ for each field.

There was strong HCO$^+$ emission at velocities near -140 km s$^{-1}$ in
each of the four images.
Figure 1 shows the low-resolution HCO$^+$ spectrum averaged
over areas of
all four images where the line was above the noise.  We integrated the image
cubes over the velocity range -142.5 to -125.7 km s$^{-1}$ where the
line emission was strongest, and mosaicked the four images using standard
AIPS tasks.  Figure 2 shows the resulting image.

In Fig. 2 the three `$+$s' correspond to the tip of
the northward-pointing VLA radio jet, the flat-spectrum radio core,
and the tip of the southern jet, respectively.  The HCO$^+$ emission
does not coincide with the location of \1e, but it does lie along a
ridge 15\arcsec\ to the west, oriented roughly parallel to the jets.
The high resolution HCO$^+$ spectra (not shown) indicate a slight
velocity gradient along the ridge.  As one moves south from the
northern tip of the ridge the velocity of the HCO$^+$ line
increases from -142 km s$^{-1}$ to -138 km s$^{-1}$ over a distance of
50\arcsec\ (2 pc at the Galactic center.) The southernmost part of the
ridge -- a single-contoured ``blob'' near the lowest `$+$' in Figure 2
-- is at a velocity of -130 km s$^{-1}$.  It may therefore be a
cloudlet which is not associated with the rest of the ridge, but
appears to be so in projection along the line of sight.

\section{Discussion}

Figure 2 clearly shows that there is no detectable HCO$^+$
emission near the radio lobes or the X-ray source.  In this
section we discuss whether the absence of HCO$^+$ emission is consistent
with \1e being the Great Annihilator.

Krolik \& Kallman (1983) have considered molecular cloud chemistry in the
presence of ionization from embedded X-ray sources.  They found that X-ray
chemistry is very much like cosmic-ray powered chemistry in that it is the
secondary electrons from X-ray ionizations which perform most of the
chemically important ionizations. Consequently, the important
factor in assessing molecular abundances is the ionization rate.  Although
they were concerned with sources of soft X-rays (primarily stellar), their
results are applicable here because the molecular cloud is optically thin at
high energies, and soft X-rays
($E\lesssim 1$ keV) will be most important for the chemistry of the cloud.

At low to moderately high ionization rates, $\xi\lesssim 10^{-15}$ s$^{-1}$,
the abundance of HCO$^+$ increases with increasing ionization rate.
However, at very high ionization rates, $\gtrsim 10^{-14}$ s$^{-1}$, the
HCO$^+$ abundance begins to decrease with increasing ionization rates
(\cite{kro83}).  Hence, a possible explanation for the lack of HCO$^+$
emission in the neighborhood of \1e is that a sufficiently high ionization
rate either destroys any existing HCO$^+$ or inhibits its formation.

We have calculated the ionization rate in the neighborhood of \1e. Our
model assumes that the emergent spectrum of \1e is that of a Comptonized disk
(Fig. 6 of Churazov, 1993) and that the source is surrounded by
atomic hydrogen of uniform density, $n_{\rm H}$.  We have assumed atomic rather
than molecular hydrogen because the photoionization cross section of
molecular hydrogen is poorly known at high energies (\cite{kro83}).
We used the approximate expressions given by
Shapiro, Lightman, \& Eardley (1976) to calculate the emergent spectrum from
the disk.

As we have already discussed, most of the ionization is due to softer
photons, $< 1$ keV, about which little or no information exists.  Hence, we
shall be forced to extrapolate this spectrum to much lower energies.
Provided that the spectrum continues to increase to lower energies, rather
than levelling off or falling, we expect our conclusions to be reasonably
robust.  Note that an implicit assumption of a Comptonized disk spectrum
model is that there exists a central source which produces a copious number
of photons well below the lower limit at which \1e has been detected,
$\approx 10$ keV.  Hence, to the extent that the Comptonized disk spectrum
is an accurate model for \1e, our ionization rates may be underestimates,
potentially by a large factor.

The ionization rate as a function of distance from the X-ray source is
shown in Fig. 3 for various ambient atomic hydrogen densities.  For a dense
cloud, $n\sim 10^4$ to $10^5$ cm$^{-3}$, the ionization rate drops
to $10^{-15}$ s$^{-1}$ about 1 pc from the X-ray source.  This is in good
accord with the projected separation we observe between \1e and the HCO$^+$
ridge, $15''\approx 0.6$ pc at the Galactic center.  Given the numerous
uncertainties in the problem we regard this agreement with our observations
as remarkably good.

These ionization rates are not sufficiently high to produce any
substantial fractional ionization.  We estimate that the gas
in the neighborhood of \1e should be neutral to a few parts in $10^4$,
assuming a nominal density and temperature of $10^5$
cm$^{-3}$ and $10^4$ K (\cite{lin88}).  This estimate is
consistent with the radio continuum observations of Mirabel et al.
(1992) and Anantharamaiah et al. (1993) who found no \ion{H}{2}
region near \1e.  Hence, this cloud could serve as the cold, dense
annihilation region for the Great Annihilator.

A more conclusive link between the molecular cloud and \1e could be obtained
by imaging the field around \1e in the lines of molecules which have enhanced
abundances in high-ionization environments.  Millimeter-wavelength
transitions of CN and HCN$^+$ are ideal for this purpose.  Krolik \& Kallman
(1983) discuss these and other molecules marked by increasing abundances
even at ionization rates as high as $10^{-13}$ s$^{-1}$.  The detection of
such molecules near the core of \1e would definitively establish the
Galactic origin of this object.

\acknowledgements

We thank Paul Goldsmith for helpful discussions.  This work was supported by
NSF Grant AST 90-16404.

\clearpage
\begin{figure}
\caption[Spectra of \1e]
{HCO$^+$ spectrum of \1e obtained
with the OVRO millimeter interferometer.}
\label{fig:spectra}
\end{figure}

\begin{figure}
\caption[Image of \1e]
{The HCO$^+$ emission integrated over the velocity range -142.5 to -125.7 km
s$^{-1}$. The crosses mark the location of the central component and the two
lobes of the radio counterpart to \1e. The image is a mosaic of four
overlapping pointing centers around \1e.  Before mosaicking, the images of
the four pointing centers were corrected for the gaussian response of the
65\arcsec\ primary beam with the result that the noise at the edge of the final
image (rms$\approx 30$ mJy/beam) was 1.5 times higher than the noise at the
center (20 mJy/beam).  The brightness contours are at
-100,-80,-60,-40,40,60,80,100,120, and 140 mJy/beam.  The deep negative
features in the map are indicative of extended emission which was ``resolved
out'' by the interferometer.  Comparing our image to the IRAM map of HCO$^+$
emission, we find that the total flux in the synthesis image is $\lesssim 20$\%
of that in the single-dish map.}
\label{fig:image}
\end{figure}

\begin{figure}
\caption[The ionization rate in the neighborhood of \1e.]
{The ionization rate as a
function of distance from \1e. The dashed horizontal line indicates the
order of magnitude ionization rate for which HCO$^+$ production is enhanced.
For a cloud with uniform density $10^5$ cm$^{-3}$ we would expect to see
an enhancement of HCO$^+$ abundance $\approx 0.7$ pc from the X-ray source,
which is in good accord with what is observed.}
\label{fig:ionize}
\end{figure}

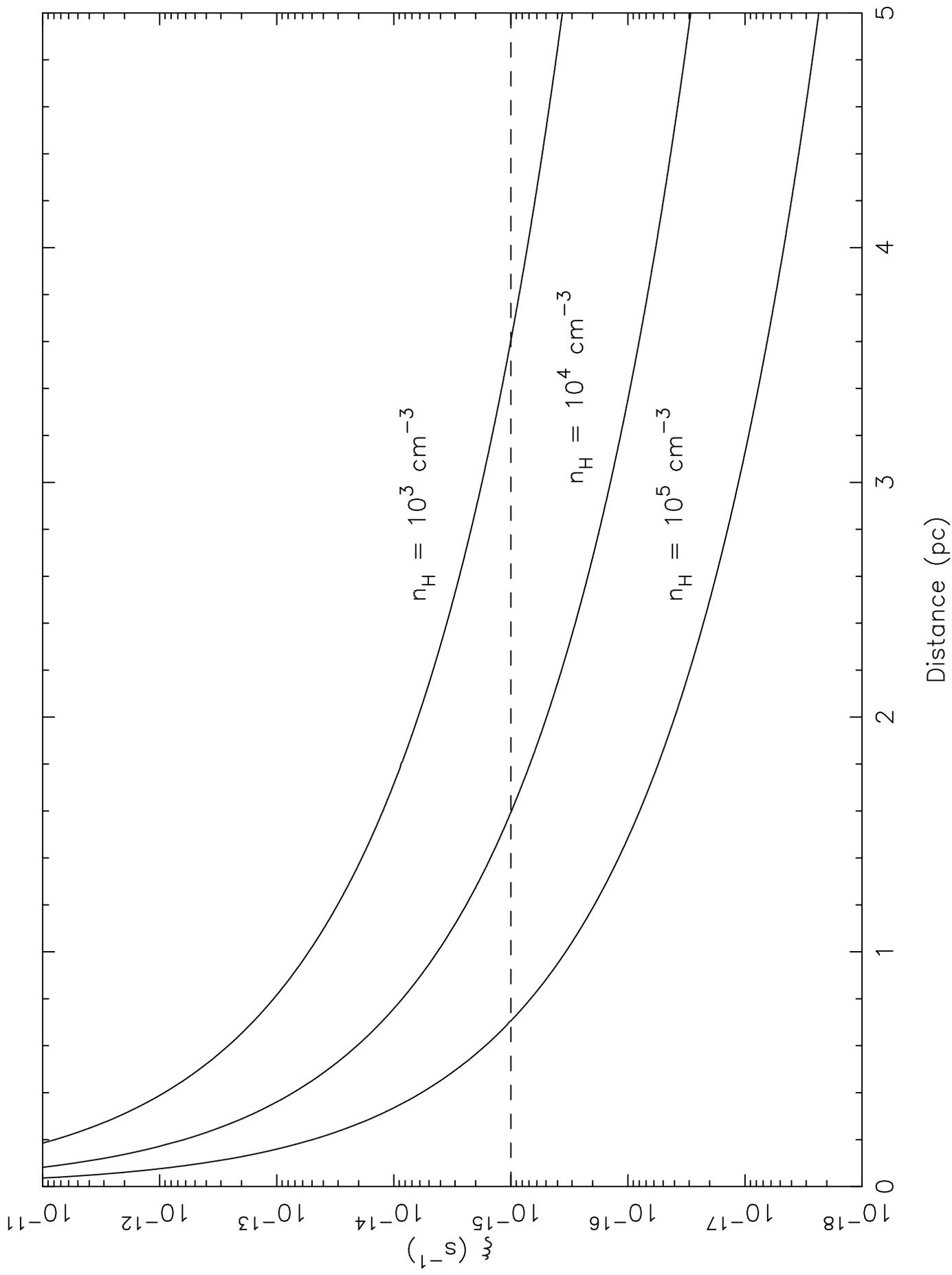
\end{document}